
\magnification=1200
\hoffset=-.1in
\voffset=-.2in

\vsize=7.5in
\hsize=5.6in
\tolerance 10000

\def\gboxit#1{\hbox{\vrule\vbox{\hrule\kern3pt\vtop
{\hbox{\kern3pt#1\kern3pt}
\kern3pt\hrule}}\vrule}}

\def\ttilde#1{\raise2ex\hbox{${\scriptscriptstyle(}\!
\sim\scriptscriptstyle{)}$}\mkern-16.5mu #1}
\def\dddots#1{\raise1ex\hbox{$^{\ldots}$}\mkern-16.5mu #1}
\def\siton#1#2{\raise1.5ex\hbox{$\scriptscriptstyle{#2}$}\mkern-17.5mu #1}
\def\pp#1#2{\raise1.5ex\hbox{${#2}$}\mkern-17mu #1}
\def\upleftarrow#1{\raise2ex\hbox{$\leftarrow$}\mkern-16.5mu #1}
\def\uprightarrow#1{\raise2ex\hbox{$\rightarrow$}\mkern-16.5mu #1}
\def\upleftrightarrow#1{\raise1.5ex\hbox{$\leftrightarrow$}\mkern-16.5mu #1}
\def\bx#1#2{\vcenter{\hrule \hbox{\vrule height #2in \kern #1in\vrule}\hrule}}

\def\squiggle#1{\lower1.5ex\hbox{$\sim$}\mkern-14mu #1}

\def\narrower{\advance\leftskip by\parindent \advance\rightskip by\parindent}

\def\onsim#1{\hbox{$ {     \lower.40ex\hbox{$\sim$}
                   \atop \raise.20ex\hbox{$#1$}
                   }     $}  }

\def\simon{\hbox{$ {     \lower.40ex\hbox{$+$}
                   \atop \raise.20ex\hbox{$\sim$}
                   }     $}  }


\def\mbox#1#2{\vcenter{\hrule width#1in\hbox{\vrule height#2in
   \hskip#1in\vrule height#2in}\hrule width#1in}}
\def\eqsquare #1:#2:{\vcenter{\hrule width#1\hbox{\vrule height#2
   \hskip#1\vrule height#2}\hrule width#1}}
\def\inbox#1#2#3{\vcenter to #2in{\vfil\hbox to #1in{$$\hfil#3\hfil$$}\vfil}}
\def\strutdepth{\dp\strutbox}
\def\marbul{\strut\vadjust{\kern-\strutdepth\specialbul}}
\def\specialbul{\vtop to \strutdepth{
    \baselineskip\strutdepth\vss\llap{$\bullet$\qquad}\null}}
\def\Bcomma{\lower6pt\hbox{$,$}}    
\def\bcomma{\lower3pt\hbox{$,$}}    

\def\sl{\scrsf}

\def\updots{\mathinner{\mskip 1mu\raise 1pt\hbox{.}
    \mskip 2mu\raise 4pt\hbox{.}\mskip 2mu
    \raise 7pt\vbox{\kern 7pt\hbox{.}}\mskip 1mu}}

\def\square{\kern1pt\vbox{\hrule height 1.2pt\hbox{\vrule width 1.2pt\hskip 3pt
   \vbox{\vskip 6pt}\hskip 3pt\vrule width 0.6pt}\hrule height 0.6pt}\kern1pt}
\def\ssquare{\kern1pt\vbox{\hrule height .6pt\hbox{\vrule width .6pt\hskip 3pt
   \vbox{\vskip 6pt}\hskip 3pt\vrule width 0.6pt}\hrule height 0.6pt}\kern1pt}
\def\lege{\hbox{$ {     \lower.40ex\hbox{$>$}
                   \atop \raise.20ex\hbox{$<$}
                   }     $}  }

\def\rege{\hbox{$ {     \lower.40ex\hbox{$<$}
                   \atop \raise.20ex\hbox{$>$}
                   }     $}  }

\def\lapp{\hbox{$ {     \lower.40ex\hbox{$<$}
                   \atop \raise.20ex\hbox{$\sim$}
                   }     $}  }
\def\rapp{\hbox{$ {     \lower.40ex\hbox{$>$}
                   \atop \raise.20ex\hbox{$\sim$}
                   }     $}  }

\def\tridots{\hbox{$ {     \lower.40ex\hbox{$.$}
                   \atop \raise.20ex\hbox{$.\,.$}
                   }     $}  }
\def\Times{\times\hskip-2.3pt{\raise.25ex\hbox{{$\scriptscriptstyle|$}}}}

\def\rightonleft{\hbox{$ {     \lower.40ex\hbox{$\longrightarrow$}
                   \atop \raise.20ex\hbox{$\longleftarrow$}
                   }     $}  }

\def\pmb#1{\setbox0=\hbox{$#1$}%
\kern-.025em\copy0\kern-\wd0
\kern.05em\copy0\kern-\wd0
\kern-.025em\raise.0433em\box0 }

\def\sqr#1#2{{\vbox{\hrule height.#2pt \hbox{\vrule width.#2pt
    height#1pt \kern#1pt \vrule width.#2pt} \hrule height.#2pt}}}

\def\inbar{\,\vrule height 1.5ex width .4pt depth0pt}
\def\IB{\relax{\rm I\kern-.18em B}}
\def\IC{\relax\hbox{$\inbar\kern-.3em{\rm C}$}}
\def\ID{\relax{\rm I\kern-.18em D}}
\def\IE{\relax{\rm I\kern-.18em E}}
\def\IF{\relax{\rm I\kern-.18em F}}
\def\IG{\relax\hbox{$\inbar\kern-.3em{\rm G}$}}
\def\IH{\relax{\rm I\kern-.18em H}}
\def\II{\relax{\rm I\kern-.18em I}}
\def\IK{\relax{\rm I\kern-.18em K}}
\def\IL{\relax{\rm I\kern-.18em L}}
\def\IM{\relax{\rm I\kern-.18em M}}
\def\IN{\relax{\rm I\kern-.18em N}}
\def\IO{\relax\hbox{$\inbar\kern-.3em{\rm O}$}}
\def\IP{\relax{\rm I\kern-.18em P}}
\def\IQ{\relax\hbox{$\inbar\kern-.3em{\rm Q}$}}
\def\IR{\relax{\rm I\kern-.18em R}}
\def\IZ{\relax\ifmmode\lrefhchoice
{\hbox{\cmss Z\kern-.4em Z}}{\hbox{\cmss Z\kern-.4em Z}}
{\lower.9pt\hbox{\cmsss Z\kern-.4em Z}}
{\lower1.2pt\hbox{\cmsss Z\kern-.4em Z}}\else{\cmss Z\kern-.4em Z}\fi}
\font\cmss=cmss10 \font\cmsss=cmss10 at 7pt
\def\IZ{\relax\ifmmode\mathchoice
{\hbox{\cmss Z\kern-.4em Z}}{\hbox{\cmss Z\kern-.4em Z}}
{\lower.9pt\hbox{\cmsss Z\kern-.4em Z}}
{\lower1.2pt\hbox{\cmsss Z\kern-.4em Z}}\else{\cmss Z\kern-.4em Z}\fi}



\font\scrsf=cmssi10                

\font\ninebf=cmbx9


%
%
\font\fivebf=cmbx5
\font\sixbf=cmbx6
\font\sevenbf=cmbx7
\font\eightbf=cmbx8
\font\ninebf=cmbx9
\font\tenbf=cmbx10

\font\bfmone=cmbx10 scaled\magstep1

\font\sevenit=cmti7
\font\eightit=cmti8
\font\nineit=cmti9
\font\tenit=cmti10

\font\itmone=cmti10 scaled\magstep1

\font\fiverm=cmr5
\font\sixrm=cmr6
\font\sevenrm=cmr7
\font\eightrm=cmr8
\font\ninerm=cmr9
\font\tenrm=cmr10

\font\rmmone=cmr10 scaled\magstep1

\def\fontone{\def\rm{\fcm0\rmmone}%
  \textfont0=\rmmone \scriptfont0=\tenrm \scriptscriptfont0=\sevenrm
  \textfont1=\itmone \scriptfont1=\tenit \scriptscriptfont1=\sevenit
  \def\it{\fcm\itfcm\itmone}%
  \textfont\itfcm=\itmone
  \def\bf{\fcm\bffcm\bfmone}%
  \textfont\bffcm=\bfmone \scriptfont\bffcm=\tenbf
   \scriptscriptfont\bffcm=\sevenbf
  \tt \ttglue=.5em plus.25em minus.15em
  \normalbaselineskip=25pt
  \let\sc=\tenrm
  \let\big=\tenbig
  \setbox\strutbox=\hbox{\vrule height10.2pt depth4.2pt width\z@}%
  \normalbaselines\rm}



\font\ninerm=cmr9
\font\eightrm=cmr8
\font\sixrm=cmr6

\font\ninei=cmmi9
\font\eighti=cmmi8
\font\sixi=cmmi6
\skewchar\ninei='177 \skewchar\eighti='177 \skewchar\sixi='177

\font\ninesy=cmsy9
\font\eightsy=cmsy8
\font\sixsy=cmsy6
\skewchar\ninesy='60 \skewchar\eightsy='60 \skewchar\sixsy='60

\font\ninebf=cmbx9
\font\eightbf=cmbx8
\font\sixbf=cmbx6

\font\ninett=cmtt9
\font\eighttt=cmtt8

\hyphenchar\tentt=-1 
\hyphenchar\ninett=-1
\hyphenchar\eighttt=-1

\font\ninesl=cmsl9
\font\eightsl=cmsl8

\font\nineit=cmti9
\font\eightit=cmti8


\newskip\ttglue
\def\tenpoint{\def\rm{\fcm0\tenrm}%
  \textfont0=\tenrm \scriptfont0=\sevenrm \scriptscriptfont0=\fiverm
  \textfont1=\teni \scriptfont1=\seveni \scriptscriptfont1=\fivei
  \textfont2=\tensy \scriptfont2=\sevensy \scriptscriptfont2=\fivesy
  \textfont3=\tenex \scriptfont3=\tenex \scriptscriptfont3=\tenex
  \def\it{\fcm\itfcm\tenit}%
  \textfont\itfcm=\tenit
  \def\sl{\fcm\slfcm\tensl}%
  \textfont\slfcm=\tensl
  \def\bf{\fcm\bffcm\tenbf}%
  \textfont\bffcm=\tenbf \scriptfont\bffcm=\sevenbf
   \scriptscriptfont\bffcm=\fivebf
  \def\tt{\fcm\ttfcm\tentt}%
  \textfont\ttfcm=\tentt
  \tt \ttglue=.5em plus.25em minus.15em
  \normalbaselineskip=16pt
  \let\sc=\eightrm
  \let\big=\tenbig
  \setbox\strutbox=\hbox{\vrule height8.5pt depth3.5pt width\z@}%
  \normalbaselines\rm}

\def\ninepoint{\def\rm{\fcm0\ninerm}%
  \textfont0=\ninerm \scriptfont0=\sixrm \scriptscriptfont0=\fiverm
  \textfont1=\ninei \scriptfont1=\sixi \scriptscriptfont1=\fivei
  \textfont2=\ninesy \scriptfont2=\sixsy \scriptscriptfont2=\fivesy
  \textfont3=\tenex \scriptfont3=\tenex \scriptscriptfont3=\tenex
  \def\it{\fcm\itfcm\nineit}%
  \textfont\itfcm=\nineit
  \def\sl{\fcm\slfcm\ninesl}%
  \textfont\slfcm=\ninesl
  \def\bf{\fcm\bffcm\ninebf}%
  \textfont\bffcm=\ninebf \scriptfont\bffcm=\sixbf
   \scriptscriptfont\bffcm=\fivebf
  \def\tt{\fcm\ttfcm\ninett}%
  \textfont\ttfcm=\ninett
  \tt \ttglue=.5em plus.25em minus.15em
  \normalbaselineskip=11pt
  \let\sc=\sevenrm
  \let\big=\ninebig
  \setbox\strutbox=\hbox{\vrule height8pt depth3pt width\z@}%
  \normalbaselines\rm}

\def\eightpoint{\def\rm{\fcm0\eightrm}%
  \textfont0=\eightrm \scriptfont0=\sixrm \scriptscriptfont0=\fiverm
  \textfont1=\eighti \scriptfont1=\sixi \scriptscriptfont1=\fivei
  \textfont2=\eightsy \scriptfont2=\sixsy \scriptscriptfont2=\fivesy
  \textfont3=\tenex \scriptfont3=\tenex \scriptscriptfont3=\tenex
  \def\it{\fcm\itfcm\eightit}%
  \textfont\itfcm=\eightit
  \def\sl{\fcm\slfcm\eightsl}%
  \textfont\slfcm=\eightsl
  \def\bf{\fcm\bffcm\eightbf}%
  \textfont\bffcm=\eightbf \scriptfont\bffcm=\sixbf
   \scriptscriptfont\bffcm=\fivebf
  \def\tt{\fcm\ttfcm\eighttt}%
  \textfont\ttfcm=\eighttt
  \tt \ttglue=.5em plus.25em minus.15em
  \normalbaselineskip=9pt
  \let\sc=\sixrm
  \let\big=\eightbig
  \setbox\strutbox=\hbox{\vrule height7pt depth2pt width\z@}%
  \normalbaselines\rm}





\baselineskip 12pt plus 1pt minus 1pt
\pageno=0
\centerline{\bf AHARONOV-BOHM SCATTERING,}
\smallskip
\centerline{{\bf CONTACT INTERACTIONS AND SCALE  INVARIANCE}\footnote{*}{This
work is supported in part by funds
provided by the U. S. Department of Energy (D.O.E.) under contract
\#DE-AC02-76ER03069.}}
\vskip 24pt
\centerline{O.~Bergman and G.~Lozano\footnote{$^\dagger$}{Financially
supported by the World Laboratory Scholarship Program.} }
\vskip 12pt
\centerline{\it Center for Theoretical Physics}
\centerline{\it Laboratory for Nuclear Science}
\centerline{\it and Department of Physics}
\centerline{\it Massachusetts Institute of Technology}
\centerline{\it Cambridge, Massachusetts\ \ 02139\ \ \ U.S.A.}
\vskip 1in
\centerline{\bf ABSTRACT}
\medskip
{\noindent\narrower
We perform a perturbative analysis of the Aharonov-Bohm problem to one
loop in a field-theoretic formulation, and show that contact interactions
are necessary for renormalizability. In general, the classical scale invariance
of this problem is broken quantum mechanically. There exists however a critical
point for which this anomaly disappears. \smallskip}
\vskip 1in
\centerline{Submitted to: {\it Physics Letters B }}
\vfill
\centerline{ Typeset in $\TeX$ by Heather Grove}
\vskip -12pt
\noindent CTP\#2182  \hfill January 1993
\eject

\vfill
\eject
\noindent{\bf I.\quad INTRODUCTION}
\medskip
\nobreak

The Aharonov-Bohm (AB) effect, which is essentially the scattering of charged
particles by a flux tube [1], has been one of the most studied problems of
planar physics in the last thirty years [2].  Much of the recent interest in
the subject is a consequence of the discovery that flux-charge composites
acquire fractional statistics (becoming anyons) through the AB effect [3].
Such composites can be obtained by coupling ordinary particles (bosons or
fermions) to a gauge field, whose dynamics are governed not by the Maxwell
action but by the Chern-Simons (CS) action [4].  Thus, scattering of a charged
particle from a flux tube, scattering of two anyons, and scattering of two
particles coupled to a CS gauge field are all the same problem.

In their original work, Aharonov and Bohm found an exact expression for the
scattering amplitude, which has been since rederived in different ways
[2,5,6].  Nevertheless, several attempts to obtain the result
perturbatively have failed [7].  As recognized by Corinaldesi and Rafeli, the
Born approximation misses the s-wave contribution to the scattering
amplitude in first order, while the second term in the Born series is
infinite.

Not surprisingly, similar problems with the perturbative calculation have
appeared more recently in the context of anyon physics in the so-called
bosonic end [8,9,10].  In this case the quantities of interest are the
eigenenergies and thermodynamic properties of a system of anyons. Sense is made
 of the perturbation theory by either redefining the unperturbed wavefunction
[8]
, or by solving a different but equivalent problem with a transformed
wavefunction
 and a transformed Hamiltonian [9,10].  In the latter method the transformed
Hamiltonian is actually not Hermitian, and it is interesting to note that
if this is corrected by adding to it its conjugate, a $\delta$-function
potential
 arises [10].  Though these methods reproduce to lowest order the correct
result
 for two anyons, the manipulations necessary for this appear arbitrary.

The aim of our paper is to clarify the perturbative analysis of the AB
problem, and to show that a correct treatment must include contact
interactions.  We shall do this in the framework of quantum field theory by
using the second quantized formulation of the AB effect for bosons, developed
in [11].  We shall see that the contact interaction, first considered in [12],
is necessary to ensure the
renormalizability of the theory.  In part I we shall write down the Lagrangian
density for this theory, review how it reduces to the AB problem, and show
that it is formally scale invariant.  In part II we shall perform a standard
perturbative analysis to one loop, and show that in general renormalization is
necessary, and results in the breaking of scale invariance, as occurs in a
theory with only contact interaction [13].  However, for a certain strength of
the contact interaction, whether attractive or repulsive, the theory has a
critical point [14], and scale invariance is
regained.  In the repulsive case, this value of the strength also
reproduces the AB result.  In the concluding remarks we comment on how our
analysis changes if we solve the constraints for the gauge field before doing
perturbation theory, and how it changes for fermions.

\goodbreak
\bigskip
\noindent{\bf II.\quad FIELD-THEORETICAL FORMULATION OF THE AB EFFECT}
\medskip
\nobreak

We begin by considering a system of nonrelativistic bosons in $2+1$
dimensions minimally coupled to a CS gauge field.  The Lagrangian
density then reads:
$${\cal L}={\kappa\over 2}\partial _t{\bf A}\times{\bf A}-\kappa A_0 B+\phi
^*\left(iD_t+{{\bf D}^2\over 2}\right)\phi-V[\phi,\phi ^*]\ ,
\eqno(\hbox{II.1})$$
where the covariant derivatives are
$$D_t=\partial_t + ieA_0 \eqno(\hbox{II.2a})$$
$${\bf D}=\pmb{\nabla} - ie{\bf A}\ , \eqno(\hbox{II.2b})$$
and the mass is set to unity.
This model, with the potential term $V[\phi,\phi^*]$ set
to
zero, was first considered by Hagen [11] as an example of a Galilean invariant
gauge theory.  The Lagrangian density (II.1) also coincides with the one
proposed by other authors [15] as an effective theory for the Fractional
Quantum Hall Effect.  We shall consider the potential corresponding to the
$\delta$-function, or contact, interaction:
$$V[\phi,\phi^*]={v_0\over 4}\phi^*\phi^*\phi\phi\ . \eqno(\hbox{II.3})$$  (To
study
the thermodynamic properties of the system one also includes a quadratic term,
whose coefficient is the chemical potential.  Since we are only interested in
scattering processes in the vacuum, we drop this term.)

It was shown in reference [11] that this model, {\it with $v_0=0$}, is a
field-theoretical formulation of the AB problem.  To
see this, first notice that using the Gauss Law $$\pmb{\nabla}\times{\bf
A}=-{e\over\kappa}\phi^*\phi\ , \eqno(\hbox{II.4})$$ the gauge fields can be
expressed
 in terms of
the matter fields (in the Coulomb gauge) as:  $${\bf A}({\bf r},t)=-{e\over
\kappa}\pmb{\nabla}\times\int d^2r'G({\bf r}-{\bf r'})\phi^*({\bf
r'},t)\phi({\bf
r'},t)\ , \eqno(\hbox{II.5})$$  where $G({\bf r})$ is the Green's function of
the two dimensional Laplacian
$$G({\bf r})={1\over 2\pi}\ln\vert{\bf r}\vert\ . \eqno(\hbox{II.6})$$ After
imposing
 canonical
commutation relations $$[\phi({\bf r},t), \phi^*({\bf r'},t)]=\delta({\bf
r}-{\bf r'})\ , \eqno(\hbox{II.7})$$ and identifying the Hamiltonian as
$$H=\int d^2r\left[ {1\over 2}({\bf D}\phi)^*\cdot({\bf
D}\phi)+V[\phi,\phi^*]\right]\ , \eqno(\hbox{II.8})$$ one can show that the
wave
function for the two particle sector satisfies the following Schr\"odinger
equation
$$E\psi({\bf r}_1,{\bf r}_2)=\left[-{1\over 2}\sum^2_{i\neq
j}\left(\pmb{\nabla}_i-{ie^2\over\kappa}\pmb{\nabla}_i\times G({\bf r}_i-{\bf
r}_j)\right)^2+{v_0\over 2}\delta({\bf r}_1-{\bf r}_2)\right]\psi({\bf
r}_1,{\bf r}_2)\ . \eqno(\hbox{II.9})$$

For $v_0=0$ this becomes in the center of mass (c.m.) frame
$$\left[{1\over r}{\partial\over\partial r}r{\partial\over\partial
r}-{(m+\alpha
)^2\over
r^2}\right]\Psi(r)=-E\Psi(r)\ , \eqno(\hbox{II.10})$$ where
$$\psi({\bf r})=e^{im\theta}\Psi(r) \eqno(\hbox{II.11})$$ and
$$\alpha={e^2\over 2\pi\kappa}\ , \eqno(\hbox{II.12})$$
which is exactly the equation studied by Aharonov and Bohm.
The scattering amplitude can be calculated exactly, and for $\vert\alpha\vert<1
$ is given by
$$f(k,\theta)=\left(2\pi ik\right)^{-1/2}\sin\pi\alpha\left[\cot{\theta\over
2}-i\,\hbox{sgn}\, (\alpha)\right]\ ,\ \theta\neq 0\ . \eqno(\hbox{II.13})$$
For identical particles one must add to this the amplitude with the final
states exchanged, which is gotten by letting $\theta\rightarrow\theta - \pi$ in
(II.13), resulting in the amplitude
$$f(k,\theta)=\bigl({2\over{\pi
ik}}\bigr)^{1/2}\sin\pi\alpha\left[\cot{\theta}-i\,\hbox{
sgn}\, (\alpha)\right]\ ,\ \theta\neq 0,\pi\ . \eqno(\hbox{II.14})$$

The dependence of the amplitude on $k$ is only through the kinematical factor
$(2\pi k)^{-1/2}$, which is a sign of {\it scale invariance.\/}  This can be
seen from the fact that the parameter $\alpha$ is dimensionless.  Note also
that the inclusion of the $\delta$-function potential would not modify this
property because $v_0$ is also dimensionless.

Scale invariance of the system can also be seen in the field theory.  As shown
by Jackiw and Pi [12], under a dilation $$\delta t=2t\ \ \ \ \ \ \delta{\bf
r}={
\bf
r}\ , \eqno(\hbox{II.15})$$ the following infinitesimal transformation of the
fields
$$\eqalignno{
\delta\phi &=-[1+{\bf r}\cdot{\bf D}+2tD_t]\phi &(\hbox{II.16a})\cr
\delta{\bf A} &=[{\bf r}\times\hat zB+2t{\bf E}] &(\hbox{II.16b})\cr
\delta A_0 &={\bf r}\cdot{\bf E}\ , &(\hbox{II.16c})\cr}$$
where ${\bf
E}\equiv-\pmb{\nabla}A_0-\partial_t{\bf A}$ and
$B\equiv\pmb{\nabla}\times{\bf A}$, leaves the action invariant.

Similarly, under a special conformal transformation $$\delta t=-t^2\ \ \ \ \ \
\delta{\bf r}=-t{\bf r}\ , \eqno(\hbox{II.17})$$ the following infinitesimal
transformation
$$\eqalignno{
\delta\phi &=\left[t-{i\over 2}r^2+t{\bf r}\cdot{\bf D}+t^2D_t\right]\phi
&(\hbox{II.18a})\cr
\delta{\bf A} &=-[t{\bf r}\times\hat zB+t^2{\bf E}] &(\hbox{II.18b})\cr
\delta A_0 &=-t{\bf r}\cdot{\bf E}\ ,&(\hbox{II.18c}) \cr}$$ also leaves the
action
 invariant.

One can prove that the generators of the above two transformations together
with the Hamiltonian form a group SO(2,1).  Of course invariance of the
action does not always guarantee an invariant result, because symmetries
can be broken quantum mechanically by {\it anomalies.\/}  This is indeed what
happens if we set $e=0$ in (II.1), and consider the bosons interacting only
among themselves via the $\delta$-function potential.
The classical scale invariance of the $\delta$-function interaction is broken
by quantum effects [13,16].  Moreover, we shall show in the next section that
this is also the case when one considers (II.1) without the contact
interaction, and to obtain a scale invariant result requires a fine tuning
between $e$ and $v_0$.

\goodbreak
\bigskip
\noindent{\bf III.\quad PERTURBATION THEORY}
\medskip
\nobreak

Let us begin by reviewing the problems with the perturbative treatment of the
AB effect.  If one applies the Born approximation to the Hamiltonian
considered by Aharonov and Bohm, (II.10), instead of
obtaining $$f(k,\theta)=\alpha\left({2\pi\over ik}\right)^{1/2}\left[
\cot{\theta}-i\,\hbox{sgn}\, (\alpha)\right]+{\cal O}(\alpha^3)\ ,\ \theta\neq
0,\pi\ ,
 \eqno(\hbox{III.1})$$
one obtains the
incorrect result [7] $$f(k,\theta)=\alpha\left({2\pi\over
ik}\right)^{1/2}\!\cot{\theta}+{\cal O}(\alpha^2)\ ,\ \theta\neq 0,\pi\ ,
\eqno(\hbox{III
.2})$$
in which the non-analytic part is missing. In addition, the next order
approximation
, ${\cal O}(\alpha^2)$, diverges.

Corinaldesi and Rafeli [7] noticed that the problem can be traced to the
s-wave contribution to the scattering amplitude, which is missing in (III.2).
As can be seen in (II.10), for the s-wave ($m=0$) the perturbation is of order
${\alpha}^2$, and when one looks at the integral equation satisfied by the
solution one encounters logarithmic divergences.  This is due to the singular
nature of the potential and the nonvanishing behavior of the unperturbed
s-waves at the origin.

As we shall see, these inconsistencies are resolved by the introduction of the
contact interaction, which has been ignored in all the perturbative treatments
up until now.  In Aharonov and Bohm's exact treatment such a contact term
could be ignored since they imposed the boundary condition
$$\Psi(0)=0\eqno(\hbox{III.3})$$ for the {\it exact} wavefunction.  Such a
boundary condition
can not however be imposed on the unperturbed wavefunction in a perturbative
treatment.

We analyze the perturbative problem from a field-theoretical point of view,
since it is a more familiar setting for dealing with divergences,
renormalization and scale symmetry breaking.  The quantity we wish to study is
the four point function associated with the scattering of two identical
particles in the c.m. frame.

The nonrelativistic bosonic propagator is given in momentum space by
$$D(k)={1\over k_0 -{1\over 2}{\bf k}^2+i\epsilon}\ , \eqno(\hbox{III.4})
$$ where
we are using the relativistic notation $k=(k_0 ,{\bf k})$.  Note that
the denominator is linear in the energy, resulting in propagation which is
only forward in time.

As the gauge field is completely constrained by the equations of motion
(II.4), there are no real gauge particles.  Nevertheless, it can still be
treated dynamically in internal lines.  In order to get the gauge propagator
we add to the Lagrangian density the following Galilean invariant gauge fixing
term  $${\cal L}_{GF}={1\over \xi}(\pmb{\nabla}\cdot{\bf A})^2\ ,
\eqno(\hbox{III.5})$$ and then take the limit $\xi\rightarrow 0$ in the
propagator.
The only nonvanishing components of the propagator are given in momentum space
by
$$D_{i0}(k)=-D_{0i}(k)={i{\epsilon}^{ij}k_j\over \kappa\vert{\bf
k}\vert^2}\ . \eqno(\hbox{III.6})$$  Note that since there is no $k_0$
dependence, the coordinate space propagator is instantaneous in time, which is
another sign that the associated particle is only virtual.

As shown in fig.1, there are three vertices coming from the covariant
derivatives $$\eqalignno{
\Gamma_0 &=-ie &(\hbox{III.7a}) \cr
\Gamma_i &={ie\over 2}(p_i+p_i') &(\hbox{III.7b}) \cr
\Gamma_{ij} &=-ie^2\delta_{ij}\ , &(\hbox{III.7c}) \cr}$$
and one from the contact
interaction
$$\Gamma_{v_0}=-iv_0\ . \eqno(\hbox{III.7d})$$ The relevant graphs for the
tree level scattering amplitude are depicted in fig.2.  The amplitude of the
one-gauge-particle-exchange graph is given by
$$\eqalignno{
A^{(0)}_{\rm exc} &={e^2\over \kappa}\sin \theta \left[{1\over
1-\cos\theta}-{1\over 1+\cos\theta}\right] &(\hbox{III.8}) \cr
 &={2e^2\over \kappa}\cot\theta\ . &(\hbox{III.9}) \cr}$$  The second
term in (III.8)
corresponds to the crossed graph, where the final particle states are
exchanged.  The remaining graph is just the contribution of the contact
interaction.  The full amplitude to tree level is then
$$A^{(0)}={2e^2\over \kappa}\cot\theta-iv_0\ . \eqno(\hbox{III.10})$$  The
graphs
contributing to the one-loop scattering amplitude are shown in fig.3.
All other possible one-loop graphs vanish. As happens
in many field theories, we expect an infinite contribution that should be
regularized.  Taking into account that the scale dimensions (under the scale
transformation
(II.14)) of $k_0$ and ${\bf k
}^2$ are equal, naive power counting shows that all the graphs are potentially
divergent. The box diagram, fig.3(a), although
naively logarithmically divergent, is finite.  Its contribution in the c.m.
frame reads, after performing the $k_0$ integration,
$$\eqalign{
A^{(1)}_{\rm box} &={4ie^4\over {\kappa}^2}\int{d^2k\over (2\pi)^2}\left[{({\bf
k}\times{\bf p})({\bf k}\times{\bf p'})\over ({\bf k}+{\bf
p})^2({\bf k}+{\bf p'})^2({\bf k}^2-{\bf p}^2+i\epsilon)}\; + \; {\bf
p'}\rightarrow -{\bf p'}\right] \cr
 &=-{ie^4\over 2\pi{\kappa}^2}\bigl[\ln(\vert 2\sin\theta\vert)+i\pi\bigr]\ ,
\cr} \eqno(\hbox{III.11})$$ where ${\bf p}$ is the relative
incident momentum in
the c.m., ${\bf p'}$ is the relative scattered momentum, and $\theta$ is the
angle between them, so
$$\eqalign{\vert{\bf p}\vert & = \vert{\bf p'}\vert\ , \cr
           {\bf p}\cdot{\bf p'} & = p^2 \cos\theta\ . \cr}
\eqno(\hbox{III.12})$$

The triangle diagram, fig.3(b), gives the following integral $$A^{(1)}_{\rm
tri}={-ie^4\over{\kappa}^2}\int{d^2k\over (2\pi)^2}\left[{{\bf
k}\cdot({\bf k}-{\bf p}+{\bf p'})\over {\bf k}^2({\bf k}-{\bf
p}+{\bf p'})^2}\; + \; {\bf p'}\rightarrow -{\bf p'}\right]\ .
\eqno(\hbox{III.13})
$$
This integral is logarithmically divergent.  To regularize it, we impose an
ultra-violet cutoff $\Lambda$ to obtain
$$A^{(1)}_{\rm tri}({\bf p},{\bf p'})={-ie^4\over
2\pi{\kappa}^2}\ln{{\Lambda}^2\over 2p^2\vert\sin\theta\vert}\ . \eqno(\hbox{
III.14})$$

This divergence is nothing more than the divergent contribution of the s-wave
that we mentioned before.  In fact, this diagram corresponds to the
${\alpha}^2\over r^2$ term in (II.10), whereas the $2m\alpha\over r^2$ term
corresponds to the single gauge particle exchange diagram, fig.2(a).  The
presence of this divergence in the four-point function shows that without the
contact term the theory is not renormalizable.

The contribution of the bubble diagram, fig.3(c), is known to be
logarithmically
divergent [13], and is given by
$$\eqalign{
A^{(1)}_{v_0}({\bf p},{\bf p'}) &={-iv_{0}^2\over 2}\int{d^2k\over
(2\pi)^2}{1\over p^2-k^2+i\epsilon} \cr
 &={iv^2_0\over8\pi}\left[\ln{{\Lambda}^2\over p^2}+i\pi \right]\, ,
\cr} \eqno(\hbox{III.15})$$ and the total one-loop scattering amplitude is
given
 by
$$A^{(1)}({\bf p},{\bf p'})={i\over 8\pi}\left[v^2_0-{4e^4\over
{\kappa}^2}\right]\left[\ln{{\Lambda}^2\over p^2}+i\pi\right]\ . \eqno(\hbox{
III.16})$$
Renormalization of this amplitude is carried out by redefining the coupling
constant $v_0\,$:
$$\eqalign{
v_0 &=v+\delta v \cr
\delta v &={1\over 4\pi}\left(v^2-{4e^4\over
{\kappa}^2}\right)\ln{\Lambda\over \mu}+{\cal O}(v^3,e^6)\ , \cr} \eqno(\hbox{
III.17})$$ and
the total renormalized amplitude is given by
$$A({\bf p},{\bf p'},\mu)={2e^2\over \kappa}\cot \theta -iv+{i\over
8\pi}\left(v^2-{4e^4\over {\kappa}^2}\right)\left(\ln{{\mu}^2\over
p^2}+i\pi\right)\ . \eqno(\hbox{III.18})$$

This amplitude is not scale invariant, as can be seen by the presence of the
arbitrary mass scale $\mu$.  A related result was obtained previously by
Manuel and Tarrach using a first quantized approach to studying contact
interactions of anyons [17].

We see however that at the critical point
$$v=\pm{2e^2\over \vert\kappa\vert}\ , \eqno(\hbox{III.19})$$ the scale
dependent term vanishes, restoring scale invariance to the solution.  Upon
multiplication by the kinematic factor $(2\pi p)^{-1/2}\,$, the scattering
amplitude then becomes
$$f(p,\theta)=\alpha\left({2\pi\over ip}\right)^{1/2}[\cot\theta\mp
i\,\hbox{sgn}\, (\alpha)]\ ,\  \theta\neq 0,\pi\ , \eqno(\hbox{III.20})$$
where $\alpha$ was defined in
(II.12).  Choosing the upper sign, corresponding to a {\it repulsive} contact
interaction, reproduces the AB result with the proper modification for
identical
particles (II.14). A contact interaction of such strength has also been
considered by Ezawa and Iwazaki [18].

\goodbreak
\bigskip
\noindent{\bf IV. \quad CONCLUDING REMARKS}
\medskip
\nobreak

In this final section we would like to address two additional points.
Namely, how the analysis of the previous section changes if one treats the
gauge field not as a dynamical variable, but as a function of the scalar field
given by (II.5), and why in the fermion case scale invariance is automatic.

Substituting equation (II.5) into our Lagrangian (II.1) results in a field
theory of
scalars with local and non-local interactions.  It is not difficult to see that
the term $$\eqalign{
{\cal L}_1 &=-{ie\over 2}{\bf
A}\cdot\phi^*\upleftrightarrow{\pmb {\nabla}}\phi \cr
 &={ie^2\over 2\kappa}\int d^2r'\left[{\pmb {\nabla}}\times G({\bf r}-{\bf
r'})\right]\phi^*({\bf r'},t)\phi({\bf r'},t)\cdot\phi^*({\bf
r},t)\upleftrightarrow{\pmb{\nabla}}\phi({\bf r},t)\ , \cr}
\eqno(\hbox{IV.1})$$
plays the
role of the single gauge particle exchange diagram.  The divergent
contribution of the triangle diagram appears in this scheme from the
term $${\cal L}_2=-{e^2\over 2}A^2\phi^*\phi\ , \eqno(\hbox{IV.2})$$  when one
contracts a scalar
field from one of the gauge fields with a conjugate scalar field from the
other.

The fermionic case can be analyzed along similar lines as the bosonic case,
except that the result is {\it automatically} scale invariant.  The fermionic
Lagrangian does not contain the contact term (II.3), but instead contains the
Pauli interaction term $${\cal L}_P={e\over 2}sB\psi^*\psi\ ,
\eqno(\hbox{IV.3})$$
where
$s$ is the spin projection.  The presence of this new vertex gives rise
to a new single-gauge-particle-exchange diagram, which plays the role of the
contact interaction diagram.  Unlike in the bosonic theory, the
contribution of this diagram is fixed by the strength of the Pauli
interaction, which corresponds to the critical point
(III.19).  Scale invariance is then automatic.

Note also that, in the attractive case, this value of $v$ renders the equations
 of motion self dual [12].  It is also the value for which the system admits an
 $N=2$ supersymmetric extension [19].

To summarize, we have shown that the Lagrangian (II.1) corresponds to a theory
that in general breaks scale invariance quantum mechanically, and corresponds
to a field theoretical formulation of the AB effect only when a special
relation between the coupling constants is satisfied, for which scale
invariance is preserved.

\goodbreak
\bigskip
\noindent {\bf ACKNOWLEDGEMENTS}
\medskip
The authors would like to acknowledge Professor Roman Jackiw for suggesting
this problem, and for numerous stimulating discussions. We also thank G.
Amelino-Camelia for useful discussions. G.L. thanks J. Negele and the CTP
for hospitality.
\vfil
\eject
\centerline{\bf REFERENCES}
\medskip
\item{[1]} Y.~Aharonov and D.~Bohm, Phys.~Rev.~{\bf 115}, 485 (1959)
\medskip
\item{[2]} For a recent review see S.~Ruijsenaars, Ann.~Phys.~(NY) {\bf 146}, 1
(1983)
\medskip
\item{[3]} F.~Wilczek, Phys.~Rev.~Lett.~{\bf 48}, 1144 (1982)
\medskip
\item{[4]} C.R.~Hagen, Ann.~Phys. (NY) {\bf 157}, 342 (1984);
\medskip
\item{} D.~Arovas, J.~Schrieffer, F.~Wilczek and A.~Zee, Nucl.~Phys.~{\bf
B251}, 117 (1985)
\medskip
\item{[5]} C.R.~Hagen,  Phys.~Rev.~D {\bf 41}, 2015 (1990)
\medskip
\item{[6]} R.~Jackiw, Ann.~Phys.~(NY) {\bf 201}, 83 (1990)
\medskip
\item{[7]} E.L.~Feinberg, Soviet Phys.~Usp.~{\bf 5}, 753 (1963);
\medskip
\item{} E.~Corinaldesi and F.~Rafeli, Amer.~J.~Phys.~{\bf 46}, 1185 (1978)
\medskip
\item{} K.M.~Purcell and W.C.~Henneberger, Amer.~J.~Phys.~{\bf 46}, 1255
(1978)
\medskip
\item{[8]} C.~Chou, Phys.~Rev.~D {\bf 44}, 2533 (1991); D {\bf 45}, 1433
(1993) (E)
\medskip
\item{} C.~Chou, L.~Hua and G.~Amelino-Camelia, Phys.~Lett.~B {\bf 286}, 329
(1992)
\medskip
\item{[9]} J.~McCabe and S.~Ouvry, Phys.~Lett.~B {\bf 260}, 113 (1991)
\medskip
\item{[10]} D.~Sen, Nucl.~Phys.~B {\bf 360}, 397 (1991)
\medskip
\item{[11]} C.R.~Hagen, Phys.~Rev.~D {\bf 31}, 848 (1985)
\medskip
\item{[12]} R.~Jackiw and S.Y.~Pi, Phys.~Rev.~D {\bf 42}, 3500 (1990)
\medskip
\item{[13]} O.~Bergman, Phys.~Rev.~D {\bf 46}, 5474 (1992)
\medskip
\item{[14]} G.~Lozano, Phys.~Lett.~B {\bf 283}, 70 (1992)
\medskip
\item{[15]} See for instance E.~Fradkin, in {\it Field theories of condensed
matter systems}, (Addison-Wesley, 1991).
\medskip
\item{[16]} R.~Jackiw, in M.A.B.~B\'eg Memorial Volume, edited by A.~Ali and
P.~Hoodbhoy (World Scientific, Singapore, 1991)
\medskip
\item{[17]} C.~Manuel and R.~Tarrach, Phys.~Lett.~B {\bf 268}, 222 (1991)
\medskip
\item{[18]} Z.F.~Ezawa and A.~Iwazaki, Tohoku Univ. preprint TU-375 (1991)
\medskip
\item{[19]} M.~Leblanc, G.~Lozano and H.~Min, Ann.~Phys.~(NY) {\bf 219}, 328
(1992)

\par
\vfill
\end